\documentclass[12pt,showpacs,preprintnumbers,amsmath,amssymb]{revtex4}
\usepackage{graphicx}
\begin{document}

\newcommand{\pderiv}[2]{\frac{\partial #1}{\partial #2}}
\newcommand{\deriv}[2]{\frac{d #1}{d #2}}

\title{Kirchhoff voltage law corrected for radiating circuits}

\vskip \baselineskip

\author{Vitor Lara}

\author{Kaled Dechoum}

\affiliation{Instituto de F\'{\i}sica - Universidade Federal Fluminense \\
Av. Litor\^anea s/n \\
24210-340 \hspace{0.5cm} Niter\'oi - RJ \hspace{0.5cm} Brazil}

\date{\today}

\begin{abstract}

When a circular loop composed by a $RLC$ is put to oscillate, the oscillation  will eventually vanish
in an exponentially decaying current, even considering superconducting wires, 
due to the emission of electric and magnetic dipole radiation. 
In this work we propose a modification on the Kirchhoff voltage law by adding the radiative contributions to the energy loss as 
an effective resistance, whose value is relatively small when compared to typical resistance value, 
but fundamental to describe correctly real circuits. 
We have also analysed the change in the pattern of the radiation 
spectra emitted by the circuit as we vary both the effective and electrical resistance.

\end{abstract}
\maketitle

\vskip \baselineskip

\section{Introduction}

The Kirchhoff law of voltage is a simple yet powerful rule that is presented in virtually all Electromagnetism books. 
This law is a direct application of the energy conservation, yet there are other assumptions that are not normally discussed. 
Most of the literature argue that the typical size of circuit must be much smaller than the wavelength of the 
emitted radiation~\cite{zozaya}, as a condition for the validity of this law but some other assumptions 
have to be taken into account. 

One of them is based on Faraday's law that states that the electromotive force, defined as the line
integral of the electric field vector around the circuit closed loop,
is proportional to the time rate of change of the magnetic flux through the loop, and can be written 
as the following:
\begin{equation}
{\cal{E}} = \oint \vec{E} \cdot d\vec{l} = - \frac{d}{dt} \iint \vec{B} \cdot d\vec{s} \,,
\end{equation}
and if the current
varies in the time the electromotive force is not zero, leading to an additional term in the Kirchhoff law. 
This can be circumvented by assuming the area of the circuit, as well as the variation of the magnetic field, 
are small enough or by assuming that the current is slowly varying~\cite{zozaya}.
The second assumption, not normally stated but related to the first one, assumes that the energy 
radiated by the circuit is negligible when compared with other energy scales involved in the problem.

\begin{figure}[!htb]
\begin{center}
\vspace{0.2cm}
\includegraphics[scale=0.3]{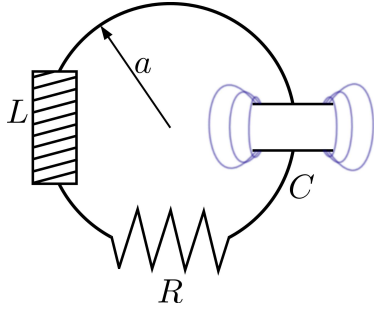}
\end{center}
\caption{A circular loop with radius $a$ composed by a $RLC$ circuit showing the capacitor radiating as electric dipole. 
The loop and the inductor also radiate (not shown) as magnetic dipole.}
\label{fig:circular_mesh}
\end{figure}

In this work we intend to overcome these hypotheses and present a simple model that hopefully clarifies the situation to some extent. 
We consider a simple $LC$ circuit. If we assume ideal capacitor, inductor and wires in the classical description, 
one obtains a perpetually oscillating current. However, intuitively we known that this result is not very realistic. 
If we consider a circular (or a square) mesh [see Fig. (\ref{fig:circular_mesh})], on the other hand, we could add into the 
recipe a term related to the magnetic and electric dipole radiation, given by the Larmor formula~\cite{griffiths}. 
In doing so, we obtain an effective resistance, which depends basically of the geometry of the mesh, of its natural 
frequency and the inductor geometry. The overall electric effective resistance is related to all sort of loss of energy, 
by the mechanical work done over the current carrying electrons and by the energy that flows out in an 
irreversible way in the form of radiation, as stated by the Poynting's theorem.

We have analysed elsewhere~\cite{RBEF} the problem of the two capacitors, the discharge of a capacitor 
associated with an identical and initially discharged, where the equilibrium state as well as the 
energy dissipated in the process are always the same, regardless the presence of a resistor in the system.
We have made clear how this system reaches equilibrium, and  
we have discussed the dissipation mechanisms and the role played by both the 
Joule effect and radiation. In this sense the present paper generalize the former and goes beyond in 
order to complement the Kirchhoff's circuit law.

What motivate us to write this article is the fact that most of the students do not well understand the fact 
that when the electric circuit radiates the traditional equation of the RLC circuit is no longer valid and  must undergo correction. 
Here we try to give a direction of how to understand the problem and how to approach it, 
giving a support for students and teachers in this subject that can be quite tricky~\cite{jackson}.

\section{Usual description of the $RLC$ circuit}

The Kirchhoff's circuit law states that for a structure such as a mesh or a circuit loop, starting from a 
point and adding voltages difference across the circuit elements, after describing a complete loop the same 
potential as that of the starting point is obtained, and its mathematical formulation is written as
\begin{equation}
\sum_{i=0}^{n} V_{i} = 0\,,
\end{equation}
where $V_{i}$ is the potential difference between the terminals of an element in the circuit, and $n$ the number of elements.

This result can be derived from the general energy balance. 
We can think in terms of the Poynting's theorem, related to the conservation of energy in the 
system formed by the electromagnetic field and matter. 
In its integral form this theorem can be written as~\cite{griffiths}
\begin{equation}
\int_{V} \vec{J} \cdot \vec{E}\,dV + \oint_{A} \vec{S} \cdot d \vec{A} = -\frac{d}{dt}
\int_{V}\frac{1}{2}(\vec{E}\cdot\vec{D} + \vec{H}\cdot\vec{B})\,dV  \, ,
\label{teoremapoynting}
\end{equation}
\noindent
where $\vec{J}$ is the current density vector, $\vec{E}$ and $\vec{B}$, as stated before, the electric and magnetic fields, 
$\vec{S}$ is the Poynting's vector, and  $\vec{D}$ and $\vec{H}$ are, respectively, 
the electric displacement and the auxiliar magnetic field vector.

This equation states that the time variation of electromagnetic energy in the field, 
the right hand side, is due to the work carried out on the charges, 
represented here by the Joule effect and by the work done over the charges by the battery, 
the first term in the left-hand side of equation (\ref{teoremapoynting}), 
and the energy that flows to the outside, represented by the surface integral of the Poynting vector, 
that represents the energy radiated by the circuit. This last term is normally neglected in 
order to calculate the current flowing in the circuit, as an approximation.

For an $RLC$ circuit forced by a battery, for example, with the assumption that the system does not radiate, 
the Poynting theorem can be cast in the form 
\begin{equation}
- R I^{2} - \frac{d }{dt} \left( \frac{L I^2}{2} \right) - \frac{d}{dt} \left(\frac{Q^2}{2C} \right) + {\cal{E}} I = 0 \,,
\end{equation}
where ${\cal{E}} I $ is the power supplied by the battery, $L I^2/2 $ is the magnetic energy stored in the inductor, 
$Q^2/2C $ is the electric energy stored in the capacitor and $ R I^{2} $ is the power dissipated in the resistor.
Using the definition of current $I = dQ/dt$ we get the standard circuit law
\begin{equation}
R I + L \frac{d I}{dt}  + \frac{Q}{C}  = {\cal{E}} \,.
\label{RLC}
\end{equation}

In the derivation of the above equations the hypothesis that the system does not radiate is implicit, and 
there is only one mechanism of energy dissipation due to the Joule effect. This assumption is not so bad when 
the electrical resistance is large enough, that is, 
much greater than the radiation loss from the circuit, but when 
we are dealing with a circuit without electric resistance such as an $LC$ circuit, 
we are induced to think that the current never stops, even without a battery source. 

In order to clarify this discussion we could remember a curious fact about superconducting 
(i.e., with no Joule dissipation at all) current carrying wires, which is somewhat 
related to the issues discussed in this work. A steady electric current flowing in a superconducting 
circular loop does not radiate, thus maintaining this current eternal. One could believe that due to 
the presence of accelerated charges this system would radiate, but the fact is that stationary currents 
do not radiate (for a detailed proof for the case of a superconducting circular loop see, 
for instance, reference \cite{Mcdonald01whydoesnt}). 

However, for the case of an $LC$ circuit formed by superconducting wires we find that the current $I(t)$ 
slowly decays as the time passes, and therefore is not stationary. Even more, we find a slowly 
decaying oscillating current, in a similar way as one can see in an $RLC$ circuit with small electrical resistance. 
As we will see later, this current leads to oscillating magnetic and electric dipoles. 
As a consequence, even an ideal $LC$ circuit radiates, which means that the current eventually vanishes. 
The main question to be addressed here is what sort of changes we have to make 
in the Kirchhoff law in order to include radiation from the circuit, and derive a differential 
equation that allows us to solve and get the circuit current as a function of time.

\section{Dissipation Mechanisms}

There are some different mechanisms for dissipation that could be included in an $RLC$ circuit aside the Joule effect, 
like the magnetic and electric dipole radiation. 
The electric dipole radiation would be due to the oscillating charge in capacitor, 
and the magnetic counterpart would be due to the geometry of the mesh and to the inductor, 
both behaving like an oscillating magnetic dipole. 
These terms are very similar, as we will see on the next subsections.

\subsection{Joule Effect}

The first contribution that we could take into account is due to the Joule effect. As stated by the law that bears the 
same name, when an electric current $I$ passes through a conductor, there is an irreversible energy transfer from the 
conductor to the medium in which it is embedded, whose power is given by $P_{R}= RI^{2}$, where $R$ is the electric 
resistance of the conductor.

This contribution is well understood and established, and leads, in the case of an $RLC$ circuit, to an electric current 
that decays exponentially.

\subsection{Electric dipole radiation}

Here we go into another well understood dissipation mechanism, although not frequently included in circuit analysis. 
Considering the accumulated charge on the capacitor plates, we have an electric dipole which is given by
\begin{equation}
 \vec{p}(t) = q(t) \vec{d} \; ,
\end{equation}
\noindent
where $d$ is the distance between the plates of the capacitor (we consider a simple capacitor of parallel plates). 
As the current varies the electric dipole $p$ is also modified, and we could include a dissipated power due to this 
electric dipole for the radiation zone ($r\gg a$, where $a$ is the radius of the RLC mesh), which is given by the Larmor formula.
The total radiated power is given by \cite{griffiths}
\begin{equation}
P_{E} = \frac{\mu_{0}}{6\pi c} \Big[\ddot{\vec{p}}(t)\Big]^{2} .
\label{electric_dipole_contribution}
\end{equation}
Since $\vec{p}(t)$ varies harmonically, this quantity can be easy calculated..

\subsection{Magnetic dipole radiation}

For the magnetic dipole radiation, there are two different sources. First, we can consider that the circular loop 
constitutes a magnetic dipole $\vec{m}_{M}(t)$, which intensity is given by
\begin{equation}
m_{M} = I(t) A \; ,
\label{magnetic_dipole_mesh}
\end{equation}
\noindent
where $A = \pi a^{2}$ is the area of the circular loop.

But there is also a magnetic dipole due to the inductor, that would be given by
\begin{equation}
m_{I} = N I(t) A^{'} \; ,
\label{magnetic_dipole_inductor}
\end{equation}
\noindent
where $N$ is the number of turns of the inductor and $A^{'}$ is the area of each one of the turns. 
Depending on the ratio between $NA^{'}$ and $A$, one of the terms could be more important than the other. 
However, for both of them, we can calculate the emitted radiation power by using
Larmor formula, given in this case by \cite{griffiths}
\begin{equation}
P_{M} = \frac{\mu_{0}}{6\pi c^{3}} \big[\ddot{\vec{m}}(t)\big]^{2} \; .
\label{magnetic_dipole_contributions}
\end{equation}

With equations (\ref{electric_dipole_contribution}) and (\ref{magnetic_dipole_contributions}) we can extend the 
Kirchhoff law of voltage, which is derived in the next section. 

\section{Generalization of Kirchhoff's Law}

As previously discussed, the usual $RLC$ circuit leads to an exponentially decaying current. If, in the other hand, 
we consider an $LC$ circuit, one obtains a perpetual oscillating current. However, this last result is somewhat artificial. 
In order to clarify this question, as pointed previously, one could consider a circular loop of radius $a$ 
(and a corresponding area $A = \pi a^{2}$). Related to this circuit, the rate of energy dissipation, 
due to the electric and magnetic dipole radiation, is given by equations (\ref{electric_dipole_contribution}) 
and (\ref{magnetic_dipole_contributions}) (this latter has two terms, see Eq. (\ref{magnetic_dipole_mesh}) 
and Eq. (\ref{magnetic_dipole_inductor})),
\begin{equation}
 P_{rad} = \frac{\mu_{0}}{6\pi c^{3}} \big[\ddot{\vec{m}}_{I}(t)\big]^{2}  + \frac{\mu_{0}}{6\pi c^{3}} 
\big[\ddot{\vec{m}}_{M}(t)\big]^{2} + \frac{\mu_{0}}{6\pi c} \Big[\ddot{\vec{p}}(t)\Big]^{2} \; , 
\label{radiaton_terms}
\end{equation}
in SI units. 

If we require energy conservation again, 
by adding this time the power formula given by Eq. (\ref{radiaton_terms}) we obtain
\begin{equation}
R I^2 + \frac{d}{dt}\Big(\frac{LI^{2}}{2}\Big) + \frac{d}{dt}\Big(\frac{Q^{2}}{2C}\Big) = {\cal{E}}I - P_{rad} \,.
\end{equation}
It is important to remark that we are neglecting retardation effect of the emitted field 
over the electric current dynamics in the circuit. This is a reasonable assumption since the 
circuit is much greater than the emitted wavelength, as stated before.

If we assume that the current in the circuit still oscillate in the form $I(t) = I_{0} \, e^{-\gamma t}\cos {\omega t}$, 
where $\omega$ is the frequency of the external electromotive force, and $\gamma$ is the damping term, 
all terms in Eq. (\ref{radiaton_terms}) leads to
\begin{equation}
P_{rad} \propto \big[ \ddot{I}(t)\big]^{2} \propto I^2.
\label{eq:radiation_resistance}
\end{equation}

This means that both, the electric and magnetic terms are responsible for the circuit's radiation, and are very similar 
to the typical dissipated power due to an electric resistance. With this in mind, albeit the radiative 
terms appear to be much more complicated, from the form of the radiative power, we again obtain an exponentially 
decaying current, and we can put all types of dissipation together in just one term in the differential equation. 

Indeed, in antenna theory, this is very well known subject, since antennas are designed to radiate \cite{balanis}.  
Usually, for an efficient antenna, the radiation resistance is the most significant contribution to the overall resistance. 
Therefore, in order to correctly describe the radiating circuits, we also need to modify the Kirchhoff voltage law, 
so that we have a consistent theory.


In the next section we perform a comparison between the contributions of each of these terms. 
For now, all we can say is, although we have followed the same steps normally used on the 
derivation of the Kirchhoff Law of Voltage, if one includes the radiation terms the perpetually oscillating 
current no longer holds, giving rise to an exponential decaying current.

\section{Comparison of dissipative contributions}

To perform a comparison between each of radiative terms and the Joule effect contribution itself, 
first we need to obtain an expression for the effective resistance due to the radiation terms. 
So, we need to write the dissipated power in terms of $R_{ef}I^{2}$. The $R_{ef}$ constant would be the effective resistance.

For the Eq. (\ref{electric_dipole_contribution}), one finds that 
\begin{equation}
P_{E} = \frac{\mu_{0}}{6\pi c} \Big[ \ddot{\vec{p}}(t)\Big]^{2} = \frac{\mu_{0}}{6\pi c} d^{2} \omega^{2} I^{2} \; ,
\end{equation}
\noindent
since $\vec{p}(t) = q(t) \vec{d}$ and we have assumed that $I(t) = I_{0} \, e^{-\gamma t}\cos {\omega t}$. This leads to 
the following effective electric radiation resistance,
\begin{equation}
R_{ef}^{E} = \frac{\mu_{0}}{6\pi c} d^{2} \omega^{2} \; .
\label{effective_resistance_elec}
\end{equation}

Similarly, the Eq. (\ref{magnetic_dipole_contributions}) gives 
\begin{equation}
R_{ef}^{M} = \frac{\mu_{0} \omega^{4}}{6 \pi c^{3}}\Big[ (N A^{'})^{2} + A^{2} \Big] \; .
\label{effective_resistance_mag}
\end{equation}

As previously stated, the relative relevance of the two terms of the magnetic 
dipole radiation depends of the ratio between $N A^{'}$ and $A$. 
The ratio between $R_{ef}^{E}$ and $R_{ef}^{M}$ is given by
\begin{equation}
\frac{R_{ef}^{E}}{R_{ef}^{M}} = \Big(\frac{dc}{\omega} \Big)^{2}\frac{1}{(N A^{'})^{2} + A^{2}}
\label{ratio_elec_mag}
\end{equation} 

We can perform some estimations for the parameters that appears in Eqs. (\ref{effective_resistance_elec}) 
and (\ref{effective_resistance_mag}). Using S.I. units we can also compare their contributions with typical 
values for the electric resistance $R$, since all of then would be measured in ohms. Under the following 
values ($\mu_{0} \sim 4 \pi \times 10^{-7}$ $H m^{-1}$, $c = 3 \times 10^8 $ 
$m/s$, $d = 10^{-3}$ $m$, $C = 9 \times 10^{-13}$ $F$, $L = 100$  $\mu H$, $A^{'} = 0.002$ $m^2$ and $A = 0.05$ $m^{2}$) one finds 
\begin{equation}
R_{ef}^{E}  \sim 10^{-9}\Omega \;\; , \;\; R_{ef}^{M} \sim 10^{-7} \Omega.
\label{comparative}
\end{equation} 

As expected, both effective radiating resistances are small when compared to typical values of electric resistance. 
However, appropriate choices for the parameters would make then comparable, enabling one to observe experimentally a 
decaying current contribution due to radiation in a superconducting circuit, or even in a ``real" suitable circuit.  
This is in fact what happens in an efficient antenna. The radiation resistance are designed to be much greater than the ohmic 
resistance \cite{balanis}. An antenna has a more complicated geometry and the electric current is not uniform, 
although the radiation mechanism is essentially the same.

\section{Time scales and spectrum}

Given that the system heats and radiates it is interesting to know the spectrum associated with 
these losses, that is, we would like to calculate the 
total power dissipated by Joule effect and radiation in a given frequency range.

The power spectrum, or the dissipated power per unit frequency is given by the expression~\cite{mandel}
\begin{equation}
P(\omega) = Re\left[ (V(\omega))^{*} I(\omega)\right] =  
Re\left[ (Z(\omega)I(\omega))^{*} I(\omega)\right] = 
R_{T} I(\omega) I(-\omega) \,,
\end{equation}
where $R_{T}= R + R_{rad}$ is the total resistance present in the circuit.

\begin{figure}[!htb]
\begin{center}
\includegraphics[scale=0.3]{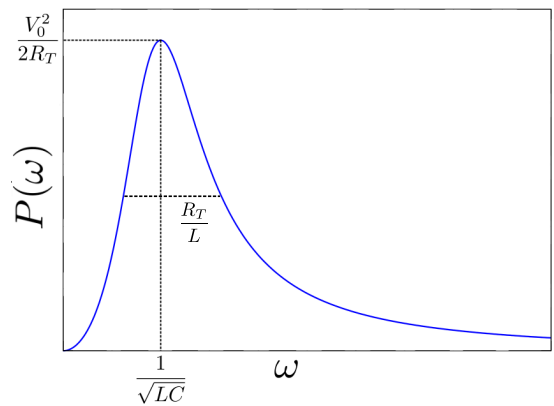}
\end{center}
\caption{The power spectrum showing its line width as function of the total resistance that include magnetic and electric radiation.}
\label{fig:P_X_w}
\end{figure}

A particular solution of equation (\ref{RLC}) can be obtained by Fourier transform method. 
Setting the battery voltage source as ${\cal{E}} (t) = V_{0} \cos \omega t$ we get the following power spectrum
\begin{equation}
P(\omega) = \frac{\frac{1}{2} R_{T} V_{0}^{2}}{R_{T}^{2} + \left( \omega L - 1/\omega C \right)^{2}} \,,
\end{equation}
where $V_{0}$ is the amplitude of battery voltage [see Fig. (\ref{fig:P_X_w})].

It is interesting to compare this spectrum for different loss mechanisms, 
when the Joule effect is more important, and when the radiation is preponderant.

The width of the distribution is determined by the resistance $R_ {T}$ that is 
present in the circuit, and it  becomes very narrow when the electrical resistance is zero, 
in this case the width of the spectral line is entirely defined by the radiation resistance, as can be seen in 
Fig. (\ref{fig:P_X_w}).

\section{Conclusions}

In this work we have made a modification in the Kirchhoff law 
of voltage, by including the Larmor power formula for both electric and magnetic dipole 
radiation in the energy conservation law, that gives rise to a differential equation for the current, 
which takes into account both Joule and radiative dissipation effects.

It is interesting to note the similarity in the expressions associated with the Joule 
dissipation, and the dipole radiation given by the Larmor formula, 
both proportional to the square of the current. It is precisely this similarity 
that allows for a simple unified analysis for dissipation. 

It is also important to have some awareness of how one can go beyond simple 
small antennas to compute the radiation resistance for circuits of arbitrary 
complexity.
If we have used another radiation system, other than a dipole that emits 
just in one frequency, we would have a different equation for the current, 
and its solution would be not so simple,
but in any case the presented change in the Kirchhoff's voltage law would still be valid. 
This is a challenge that we try to confront in the next future.

\section*{ACKNOWLEDGMENTS}

We would like to acknowledge the financial support of Brazilian agencies CAPES and CNPq, 
and for Nivaldo Lemos who proofread the manuscript.

\end{document}